# Reduction Signals Method Preserving Spatial and Temporal Capabilities

Andrea Gonzalez-Montoro, Efthymios Lamprou, Alfonso Pérez, Liczandro Hernández, Gabriel Cañizares, Marta Freire, Luis F. Vidal, Edwin J. Pincay, John Barrio, Sebastian Sánchez, Filomeno Sánchez, Jose M. Benlloch, and Antonio J. Gonzalez

*Abstract*— In gamma ray imaging, a scintillation crystal is typically used to convert the gamma radiation into visible light. Photosensors are used to transform this light into measurable signals. Several types of photosensors are currently in use depending on the application, most known are Position Sensitive Photomultiplier Tubes (PSPMT) or arrays of Silicon Photomultipliers (SiPMs). There have been investigations towards reducing the number of output signals from those photosensors in order to decrease system costs and complexity without impacting system performance. We propose here two different reduction schemes without degradation of the detector performance, keeping a good spatial, energy and timing resolution, specially well suited for monolithic scintillation crystals based detectors. We have carried out comparative results that will be shown.

*Index Terms*— Monolithic scintillators, Positron Emission Tomography, Reduction readout systems, Silicon Photomultiplier.

## I. INTRODUCTION

REDUCTION schemes reducing the number of signals but preserving system capabilities have been suggested since long time ago. The main reason has been reducing costs and complexity, related to the number of signals to post-process. In gamma ray detectors, scintillation crystals are coupled to photosensor arrays and read-out by means of frontend electronics. Scintillation crystals of a broad variety of types (LSO, LYSO, BGO, LaBr, etc) and geometries (crystal arrays or monolithic) are employed. Photosensors based on Position Sensitive Photomultiplier Tubes (PSPMT) or Silicon Photomultipliers (SiPMs) arrays are commonly used. Both have shown, more recently SiPMs, to provide comparable detector performance regarding spatial, energy and timing capabilities. The use of SiPMs are however ramping up due to its compatibility with magnetic fields, compactness and a broad offer from several providers.

A. Gonzalez-Montoro, E. Lamprou, A. Pérez, L. Hernández, G. Cañizares, M Freire, L.F. Vidal, E.J. Pincay, J. Barrio, S. Sánchez, F. Sánchez, J.M. Benlloch, and A.J. Gonzalez are with the Instituto de Instrumentación para Imagen Molecular (I3M), Centro Mixto CSIC — Universitat Politècnica de València, Camino de Vera s/n, 46022 Valencia, SPAIN (e-mail: andrea.gm@i3m.upv.es).

In the field of gamma ray detectors, initial designs made use of crystal arrays and the so-called one-to-one coupling. One-to-one coupling means individually coupling each single scintillation element to one photosensor, and consequently individual reading. This approach showed to work well when the number of elements was low, typically limited by the crystal and photosensor sizes. Small size PSPMT matching the small crystal sizes and the techniques to produce them, require the development of novel readout techniques avoiding reading individually all signals from the photosensor. One of the most known reduction read-out approaches is the Anger logic [1]. Here, a network using passive components provides four signals making it possible to determine the planar impact position of the gamma ray within the crystal, but also photon depth of interaction (DOI) in a modified version [2]. It should be pointed out that these signals are analog and are later fed up into a data acquisition system where they are digitized using Analog to Digital Converters (ADCs) or similar. Other approaches returning digital information such as free running sampling methods, or the use of Application Specific Integrated Circuits (ASICs) are also possible. Higher density photosensors, together with improved ADCs and new ASICs, suggested the possibility to increase the number of readout channels with the benefit of improving the performance of the detector block. An approach that made it possible was the use of a projection readout providing signals for each row and column of the photosensor array [3].

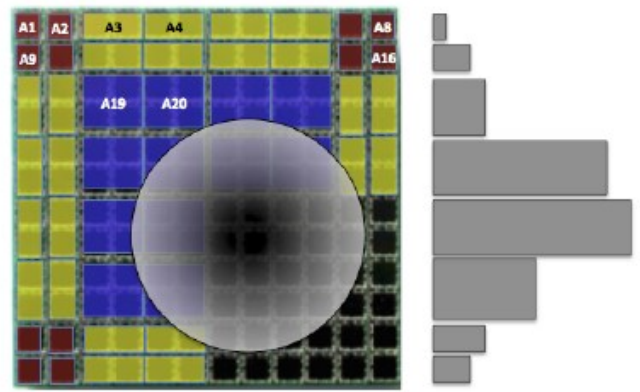

Fig. 1. Reduction schemes for a 12×12 photosensor array. The centered shadow represents a signal distribution. Left, reduction from the 144 signals to 64 by merging some at the laterals and centers. Right, projection onto the Y axis resulting in only 8 signals. Notice that 8 more signals would be projected onto the X-axis.

ASICs, especially the ones with integrated high resolution Time–To-Digital-Converters TDCs, are becoming an interesting alternative, especially when timing information is required. These devices offer a number of input channels that typically varies from 16 to up to 64. Our proposal is based on reducing the number of signals provided by a high-density photosensor array, independently of the scintillation crystal type and geometry used, but preserving the detector performance especially at the edge of the detector block, where a better sampling of the light distribution is needed. The method refers to merging signals where the low sampling of the scintillation light is possible [4][5]. Figure 1 shows a scheme of this configuration.

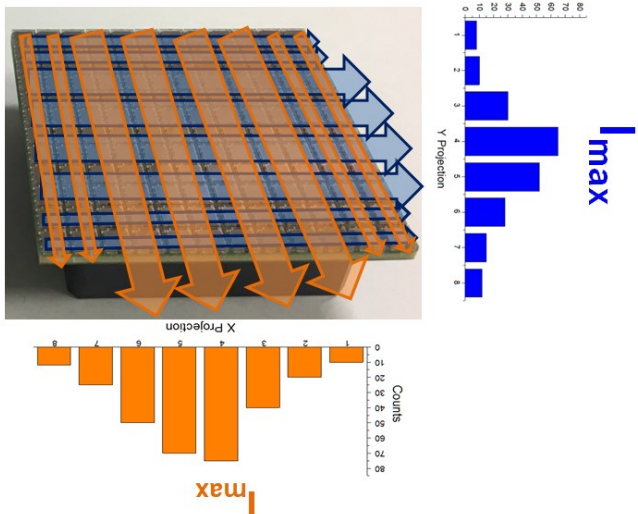

Fig. 2. Reduction readout scheme from 12×12 SiPMs to 8+8 signals. On the right (blue) projection onto the Y axis resulting in only 8 signals, and at the bottom (orange) projection onto the X axis.

## II. MATERIALS

We have based our study on a high-density array of SiPM from SensL (C-Series). The photosensor matrix is composed of 144 (12×12) individual photosensors with 3×3 mm$^2$ active area each, and a pitch of 4.2 mm. The arrays are operated at a bias voltage of 29 V, 4.5 V over the breakdown voltage [6].

We propose a two-step reduction method. On a first iteration we reduce the number of total from the original 144 down to 64. This is done by merging signals outputs; 4 to 1 at the center, and 2 to 1 at the laterals. All total output signals match well the input signals of ASICs with 64 inputs. The second step applies a reduction approach by projecting the signals onto the X and Y axes (from 64 to 16 signals), we called this method "2S". Figure 1 shows on the left the reduction by joining signals and on the right the projected signals onto one axis depicting the light distribution of an impact.

An alternative method, for which we will also show comparative data, is to first project the original 12×12 array signals onto 12+12 outputs. Thereafter, for each 12 projected signals on each axis, the 8 central ones are merged two-to-one, resulting in only 8 signals per axis, in total 16 signals. We labelled this method as "2R" [7]. Figure 2 shows a scheme of the merged signals. Both the 2S and 2R methods provide a reduction from 144 to 16 output signals. This means reducing to as less as 11% of the original signals (16÷144).

To compare the performance of the proposed reduction readout system, an additional SiPM array of 8×8 elements (SensL, J-Series) with 6×6 mm$^2$ active area each, and a pitch of 6.33 mm has been used. This array is connected to a projection circuit providing 8+8 output signals, corresponding to the number of row and column signals and without additional reduction. We referred this method as standard 8×8. The use of C- or J-series parts from SensL does not affect the results. Finally we are also comparing the performance of the 3 mentioned detector block configurations with the standard 12×12 array but connected to also a projection circuit, providing 12+12 output signals. All output signals, independently of the circuitry, are first pre-amplified and later digitized with custom ADC boards (12-bit) [4].

To evaluate the proposed reduction methods, we have made use of a monolithic LYSO crystal, with dimensions of 50×50×15 mm$^3$ with the lateral walls black painted and the entrance face coupled to a retroreflector layer that bounces back to the photosensor the scintillation light [4]. The exit face of the scintillation blocks is coupled to the photosensor using optical grease (BC630, Saint Gobain).

## III. METHODS

The performance of the proposed detector blocks was evaluated by studying their spatial and energy resolutions. Normal incidence measurements, with radioactive sources perpendicular to the entrance face of the crystal were carried out. The detector blocks under study were sequentially irradiated with an array of 11 × 11 $^{22}$Na sources, 1 mm in diameter and 1 mm height each (4.6 mm pitch), placed in front of a Tungsten collimator (24 mm thick, 1.2 mm diameter holes), which was in contact with the crystal. The reference detector was placed at a distance of 11.5 cm (crystal-to-crystal).

The centroids of the light distribution projections for both *X* and *Y* axes are calculated using the center of gravity method. The photon impact DOI is estimated by the ratio of the sum of all 8 signals (photon energy, *E*) to the maximum signal value ($I_{max}$) ($E/I_{max}$) [8].

In a further step, in order to evaluate the ability of the proposed reduction systems to characterize events closer to the crystal edge, a second set of experiments was carried out. A non-collimated small size $^{22}$Na source (0.25 mm in diameter, encapsulated in a 1 inch diameter PMMA disk), was moved in small steps of 0.5 mm across one entire axis of a crystal. This experiment was carried out for both the 2R and the 8×8 standard readout schemes. Similar values are expected for the 2S system though.

During the data processing, for all acquired data a software collimation was applied. This means that only coincidence events whose Line of Response (LOR) has a slope smaller than 2.1° from the normal, were taking into account. The selection of the angle is a tradeoff between the statistics of the analyzed measurement and its spatial resolution. Finally, an energy windowing of 15% at the 511 keV peak (434–588 keV) was also applied.

The detector spatial resolution was evaluated for three depth of interaction layers, 5 mm each one, namely DOI1 (15-10 mm, entrance), DOI2 (10-5 mm) and DOI3 (5-0 mm, face in contact with the photosensor). We calculated the centroid of each source in channels using multi-Gaussian distributions. Calibration from channels to millimeters is done by a fit to a third order polynomial. After the calibration, the detector spatial resolution is estimated as the FWHM of the multi-Gaussian fits. The energy resolution is determined as $FWHM/E_{centroid}$. Both the spatial and energy resolutions have been also studied as a function of the DOI layer.

## IV. RESULTS

Figure 3 top, shows the flood maps for the 11 × 11 collimated sources for each one of the four mentioned readout systems. From left to right standard 12×12, 2S reduction, 2R reduction and standard 8×8. As it can be appreciated in the flood maps all the 121 sources are distinguished in all cases. The central panel of figure 3 shows the spatial resolution measured as the FWHM for the central row of sources for each considered readout configuration. The bottom panel depicts the energy resolution evaluated across the diagonal of 11 sources. On average, similar results have been obtained in terms of energy and spatial resolutions for the four configurations, demonstrating that the reduction system does not worsen the detector performance.

standard 8+8. Center. Spatial resolution FWHM measured for the 1central row of sources for the 4 different readout methods. Bottom. Measured energy resolution across the diagonal of 11 sources. On average similar results are obtained for all the tested readout configurations.

Notice, that only the energy resolution for the standard 8×8 configuration shows a slightly improvement in the central region of the photosensor. Most likely this is produced because the active area in the 8×8 SiPM array is 92% whereas in the 12×12 case is only 52%. Table I summarizes the average values obtained for each case. Values are in the range of 1.8 mm and 14 % for the spatial and energy resolutions, respectively.

In a further step, we have evaluated the ability of the 2R reduction system and the standard 8+8 system to characterize events closer to the edge of the crystal, by moving the non-collimated 0.25 mm source across the X axes of the scintillator. Figure 4 shows the spatial resolution as a function of the source position. It can be appreciated that when using the standard 8×8 scheme events closer to the edge are not well resolved influencing the measured spatial resolution trend. Moreover, the measured value does not reflect the true information since the light distribution is concentrated in only one or two SiPMs. However, when using the reduction system 2R the source characterization closer to the edge is better. This is produced because the sampling is better in this area.

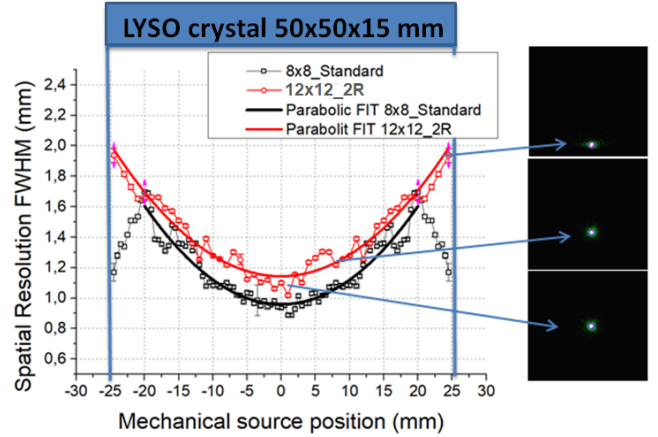

Fig. 4 Left. Measured FWHM values as a function of the known beam position across the crystal surface for the 2R reduction system and the standard 8x8 system. Right. Examples of flood map of the measurements at three different positions for the 2R system.

Finally, the detector performance has been also evaluated as a function of the DOI layer. Figure 5 shows the flood maps for the 11×11 collimated sources as a function of the DOI layer. We can observe some stronger compression effect for impacts occurring in the upper layer (DOI1), where there is a larger scintillation light truncation. For events close to the photosensor (DOI3 layer), the 2R and 2S approaches exhibit an irregular pattern of the collimated sources produced by the larger pitch generated with the proposed readout schemes. Here, the central area has a pitch that is twice the one at the edge regions. Table II and III summarize the average spatial and energy resolution values, obtained as the average of the 11 central sources for each DOI layer. For the standard 12+12 and 8+8 approaches the spatial resolution improves closer to the photosensor. However for the 2S and 2R methods, the

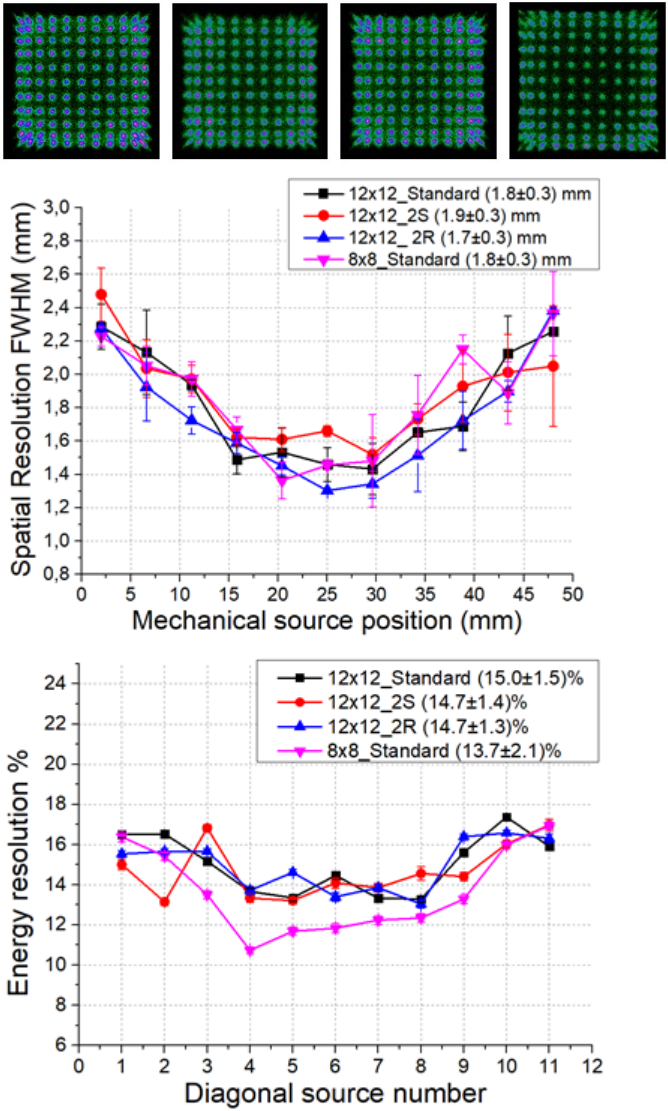

Fig. 3. Top. Flood maps of collimated sources on the 15 mm thick monolithic crystals. From left to right, standard 12+12, 2S approach, 2R approach and behavior is the opposite. This is produced by its irregular pitch size. For all cases the region closer to the entrance face of the scintillator (DOI1) depicts a worsening of the spatial resolution nearing the detector block corners due to the larger light scintillation truncation. Nevertheless, an almost homogeneous spatial resolution is obtained for the whole scintillator volume in the four cases.

TABLE I
DETECTOR BLOCK PERFORMANCE

|            | Spatial Resolution FWHM (mm) | Energy Resolution FWHM (%) |
|------------|------------------------------|----------------------------|
| 12Standard | 1.8±0.1                      | 14.9±1.8                   |
| 2S         | 1.9±0.2                      | 14.7±1.3                   |
| 2R         | 1.8±0.2                      | 14.9±1.3                   |
| 8Standard  | 1.7±0.1                      | 13.7±2.2                   |

Spatial and energy resolution measured for each readout system

TABLE II
SPATIAL RESOLUTION AS A FUNCTION OF THE DOI LAYER

|            | DOI1 (mm) | DOI2 (mm) | DOI3 (mm) |
|------------|-----------|-----------|-----------|
| 12Standard | 1.7±0.2   | 1.6±0.2   | 1.6±0.2   |
| 2S         | 1.7±0.2   | 1.7±0.1   | 1.9±0.2   |
| 2R         | 1.7±0.3   | 1.8±0.3   | 1.7±0.3   |
| 8Standard  | 1.6±0.3   | 1.5±0.2   | 1.6±0.3   |

Spatial resolution measured as the average of the X and Y FWHM source profiles

TABLE III
ENERGY RESOLUTION AS A FUNCTION OF THE DOI LAYER

|            | DOI1 (%) | DOI2 (%) | DOI3 (%) |
|------------|----------|----------|----------|
| 12Standard | 13.4±0.1 | 13.3±0.1 | 12.5±0.2 |
| 2S         | 13.2±0.2 | 12.3±0.1 | 12.4±0.2 |
| 2R         | 13.3±0.1 | 13.4±0.2 | 14.3±0.1 |
| 8Standard  | 11.9±0.1 | 12.0±0.1 | 12.3±0.2 |

Energy resolution measured as ΔE/E for the 11 sources in the diagonal

## V. DISCUSSION

We have described and tested approaches to reduce the number of signals from a SiPM array, coupled to a scintillation crystal. We have tested them using a monolithic 50×50×15 mm$^3$ block. We have validated their use when applied to gamma ray detectors, as for instance those used in PET. We do not observe significant performance degradation with regards to photon impact determination or energy resolution, especially at the detector edges. Notice that the edge sources are placed at only 0.5 mm from the crystal edge and were well resolved.

Our current research is focused on investigating the possible benefits of removing step 2 from the 2S approach, reducing signals from 144 to 64, as shown on the left panel of Figure 1. We are currently testing this readout on the ASIC TOFPET2 from PETsys. One of the advantages of this device is its temporal performance, showing Coincidence Time Resolution (CTR) as good as 200 ps (FWHM) using one-to-one coupling and 3×3×5 mm$^3$ LYSO crystals. The main aim of that study, besides spatial resolution, is to show how the timing characteristics are modified when the signals are merged.

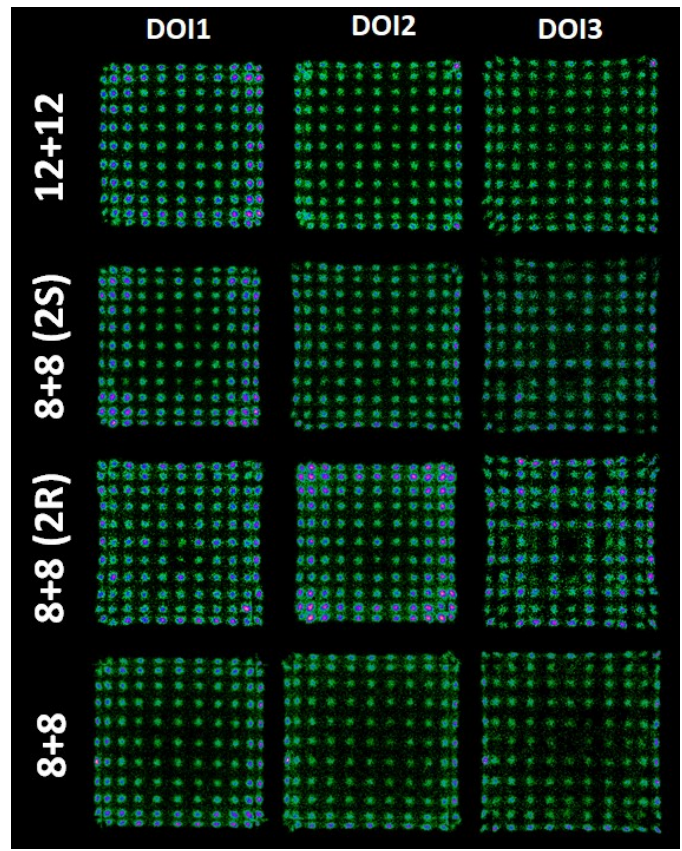

Fig. 5. Flood maps of collimated sources on the 15 mm thick monolithic crystal, for different readout approaches and as a function of DOI.